# Networking Molecular Quantum Emitters on a Single Chain
From Single to Cooperative Emitters

Jean-Baptiste Marceau[1#], Juliette Le Balle[1,2#], Christel Poujol[3], Frédéric Fossard[2], Annick Loiseau[2], Gaëlle Recher[1] and Etienne Gaufrès[1*]

[1] Laboratoire Photonique Numérique et Nanosciences, Institut d'Optique, CNRS UMR5298, Université de Bordeaux, F-33400 Talence, France

[2] Laboratoire d'Étude des Microstructures, ONERA-CNRS, UMR104, Université Paris-Saclay, BP 72, 92322 Châtillon Cedex, France

[3] Bordeaux Imaging Center, UAR 3420 CNRS – Université Bordeaux – US4 INSERM, Centre Broca Nouvelle Aquitaine, 33076 Bordeaux, France

# these authors contributed equally

*Correspondence: etienne.gaufres@cnrs.fr

**Abstract**

**Engineering light-matter interactions between multiple free-space quantum emitters is a central challenge for scalable quantum photonic technologies. In particular, accessing regimes of coherent emitter-emitter interactions, where several emitters are coupled through a shared electromagnetic environment, is essential for coherent emission and quantum functionalities. Such interactions require precise control over emitter separation and stabilization at sub-wavelength distances, a level of spatial organization that remains extremely difficult to achieve at the molecular scale in solid-state systems. Here we introduce Encoded Quantum Chains (EQC), a one-dimensional architecture in which cooperative radiative behaviour is programmed through spatial encoding of identical molecular emitters. Organic emitters and inert spacer molecules are co-encapsulated inside dielectric boron nitride nanotubes (BNNTs), enabling statistical control of intermolecular spacing from nanometres to micrometres while enforcing dipole alignment and one-dimensional confinement. Time-resolved fluorescence under ambient conditions reveals accelerated radiative decay, enhanced emission rates per emitter, and the emergence of non-mono-exponential dynamics as spacing falls below the optical wavelength, consistent with cooperative radiative states in one dimension. Bundling of EQCs enables coupling between emitters in neighbouring BNNTs, driving a dimensional crossover toward higher-dimensional delocalisation of the excitation. This modular building-block approach provides a scalable route to engineer light-matter interactions and many-body optical phenomena in confined molecular systems, opening new opportunities for distributed single-photon sources, programmable quantum emitters, and photonic architectures for quantum technologies.**

**Introduction**

Technological building blocks for on-demand single-photon and multi-photon sources are essential to quantum optics and emerging quantum technologies, enabling flying qubits for communication, computing, sensing, and metrology[1–3]. While these applications span very different regimes of light emission, they share stringent requirements on light–matter interactions at the level of individual quantum emitters[4].

Single-photon emission relies on the isolation of a single, well-defined two-levels system associated with an individual optical dipole. In contrast, the generation of correlated or entangled multi-photon states requires multiple identical emitters to radiate cooperatively into a shared electromagnetic environment. In this collective regime, coupling between emitters can give rise to super-radiant and sub-radiant states, as first described by Dicke, where the emission dynamics are governed by collective dipole synchronization rather than independent decays. Achieving such cooperative emission requires precise control over emitter spacing, dipole alignment, and coherence, a condition that in free-space systems is set by sub-wavelength length scales.

Despite sustained efforts, engineering collective radiative behaviour in solid-state platforms remains challenging. Many experimental realizations of superradiant or cooperative emission rely on cryogenic operation, lithographically defined architectures, or highly controlled but poorly scalable systems[5–7]. Elegant demonstrations of radiative coupling have been achieved in ensembles cold atoms and atoms coupled to nanophotonic waveguides, where long coherence times and controlled environments enable collective emission[8–12]. However, these approaches involve stringent experimental overheads and are difficult to translate into compact, robust, and scalable materials platforms.

At the nanoscale, collective emission has been explored in few-emitter systems, such as pairs of coherently coupled quantum emitters randomly embedded in solid organic matrices, where super-radiant and sub-radiant decay channels can be tailored[13]. While these studies provide fundamental insight into cooperative radiative dynamics, extending such control beyond two or three emitters with homogeneous, reproducible and controlled spacing as well as alignment remains a major challenge. More recently, collective-like emission signatures have been reported in self-assemblied perovskite quantum dots, where reduced radiative lifetimes from ns to ps suggest coupling between multiple emitters[14]. However, residual disorder, environmental instability, and limited tunability of the coupling strength continue to restrict systematic exploration of collective regimes.

Organic molecules constitute attractive quantum emitters due to their large oscillator strengths, chemical versatility, and room-temperature operation[15]. In rare cases, molecular self-assembly into J-aggregates can yield collective optical responses, including signatures of superradiant emission over a limited number of molecules[16–18]. Yet, the fragile nature of molecular aggregates, combined with limited control over spacing and local environment hampers reproducibility and quantitative investigation of collective photon statistics. Recent demonstration of sharp zero-phonon lines from single organic molecules hosted on boron nitride layer environments further highlight the potential of molecular emitters embedded in wide-bandgap dielectrics, while underscoring the importance of controlled nanoscale confinement and stabilization[19]. Furthermore, aggregation effects have been recently identified in molecules stacked inside boron nitride nanotubes (BNNTs), leading to collective modifications of their steady state optical spectra.[20–22]

Overall, harnessing collective radiative phenomena for quantum photonics remains limited by the lack of scalable strategies to position, align, and stabilize multiple emitters at sub-wavelength separations under ambient conditions.

Here, we introduce Encoded Quantum Chains (EQC), a one-dimensional materials platform in which collective radiative behaviour is programmed through the spatial encoding of molecular emitters. EQCs are formed by co-encapsulating organic emitters and spacer molecules inside BNNTs, which act as rigid dielectric hosts enforcing one-dimensional confinement and dipole alignment. By adjusting the relative concentration of emitters and spacers during synthesis, the average intermolecular spacing along a single chain can be statistically tuned from nanometres to micrometres.

Using this platform, we first establish a well-defined single-emitter regime at large separations when molecular emitters are widely separated and behave as isolated molecules, exhibiting fluorescence fingerprints characteristic of individual emitters at room temperature. Upon reducing the emitter spacing below the optical wavelength, we observe a pronounced acceleration of the radiative decay, the emergence of non-mono-exponential dynamics, and an enhancement of the emission rate per emitter, consistent with the formation of cooperative radiative states in a one-dimensional geometry. Finally, we show that bundling EQC@BNNT enables radiative coupling between neighbouring chains, inducing a dimensional crossover towards higher-dimensional cooperative emission. Together, these results establish EQCs as a modular and scalable platform for engineering collective light–matter interactions in confined molecular systems at room temperature.

**Main**

The concept of the Encoded Quantum Chain (EQC) is illustrated in Fig. 1a,b. It consists in positioning and stabilizing molecular quantum emitters at sub-wavelength separations along a single linear host, resembling pearls embedded within a nanoscale necklace. We exploit the crystalline, wide-bandgap (5.7 eV) one-dimensional cavity of BNNTs[20,21,23] as a rigid template to assemble EQC@BNNT structures.

The synthesis strategy relies on using the open ends of BNNTs as nanoscale gates through which functional molecules in solution enter sequentially. We hypothesize that this gating mechanism reproduces, within the confined 1D channel, the relative concentrations of molecular species present in the dilute solution, thereby yielding a densified and ordered sequence along the BNNT axis. The inner diameter of the BNNT core (1–3 nm), matching the size of the entry pore, minimizes post-encapsulation rearrangement, mixing, or sequence disruption and leaking .

Quantum emitters and spacer molecules were selected (1) to fit within the BNNT cavity, and (2) to exhibit optical resonances separated by at least 1 eV, thereby reducing spectral cross-talk and creating a longitudinal potential landscape with local minima at the emitter positions (Fig. 1b). To this end, sexithiophene (6T) and anthracene (Anth) molecules - absorbing at 2 eV and 3.5 eV, respectively - were co-solubilized with dispersed BNNTs in dimethylformamide (DMF) and refluxed at 80°C during 24h.

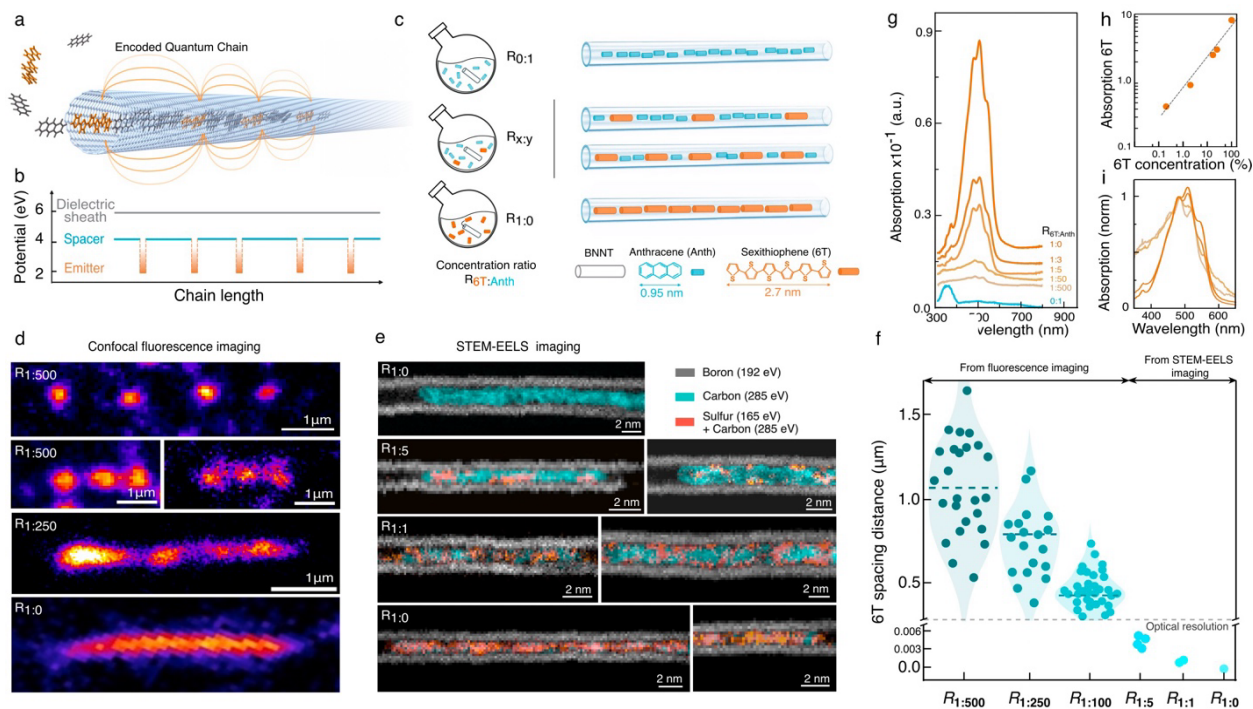

**Figure 1 – Structure and chemical composition of the Encoded Quantum Chains (EQC).**
**a** Schematic representation of EQC based on the encapsulation of sexithiophene and anthracene molecules inside a BNNT. **b** Potential function along the EQC@BNNT longitudinal axis. **c** Synthesis route for the formation of the EQC inside the BNNTs based on the liquid phase encapsulation of sexithiophene and anthracene molecules at given concentration ratio noted $R_{6T:Anth}$. **d** Confocal fluorescence imaging of EQC@BNNTs deposited on Si/SiO$_2$ substrate at an excitation wavelength of 532 nm and a collection range of 600-800 nm. **e** Chemical analysis of EQC@BNNTs: Overlay of elemental core Electron Energy Loss (EELS) images at the boron (192 eV), sulfur (165 eV), and carbon edges (285 eV), using aberration corrected Scanning Transmission Electron Microscopy ac-STEM (JEOL NeoARM) operated at 80kV, recorded from individual EQC@BNNTs suspended on holey-carbon TEM grid. **f** Violin plot representation of the statistical average spacing distance between 6T molecules among anthracene molecules inside the BNNTs as a function of $R_{6T:Anth}$. Data extracted by confocal fluorescence imaging for $R_{1:500}$ $R_{1:250}$ and $R_{1:100}$, and by STEM-EELS for $R_{1:5}$ $R_{1:1}$ and $R_{1:0}$. Colored dashed line represent the spacing distance median value for given $R_{6T:Anth}$. The optical resolution limit of our confocal set is indicated by a grey dashed line. **g** Absorption spectra of EQC@BNNT at different $R_{6T:Anth}$, redispersed in DMF solution after synthesis. **h** Integrated absorption from 400 nm to 600 nm of 6T molecules in EQC@BNNT as a function of $R_{6T:Anth}$, extracted from (g) **i** Absorption spectra normalized at 485 nm of EQC@BNNTs from $R_{1:500}$ $R_{1:50}$, $R_{1:5}$, and $R_{1:0}$.

Different 6T:anthracene concentration ratios, denoted $R_{6T:Anth}$, were prepared to tune the average intermolecular spacing within the EQCs. Mixtures ranged from pure 6T ($R_{1:0}$) to pure anthracene ($R_{0:1}$), with intermediate ratios $R_{1:1}$, $R_{1:5}$, $R_{1:50}$, $R_{1:100}$, $R_{1:250}$ and $R_{1:500}$, as schematized in Fig. 1c. Details of the EQC@BNNT synthesis are provided in the Supplementary Information.

EQC@BNNT deposited on Si/SiO$_2$ substrate by spin casting were imaged by confocal fluorescence imaging. (see supporting information file for details). Representative fluorescence images of EQC@BNNTs are presented in the Figure 1d. EQC@BNNTs synthesized at $R_{1:0}$ exhibit a continuous fluorescence signal along the nanotube axis, indicating that 6T molecules form a homogeneous, densely packed chain at the optical resolution limit (250 nm). In contrast, samples prepared at low 6T content ($R_{1:500}$ and $R_{1:250}$) display a discontinuous fluorescence profile, with discrete intensity spots distributed along the BNNT axis. These spots are attributed to 6T emitters separated by non-resonant anthracene molecules acting as spacers. The inter-emitter spacing is defined as the distance between the centroids of adjacent fluorescence spots. (Fig S1)

The distribution of inter-emitter spacing extracted from statistical ensembles of EQC@BNNTs as a function of $R_{6T:Anth}$ is shown in Fig. 1f. As expected, increasing the relative 6T concentration reduces the mean spacing distance, demonstrating that the BNNT pore statistically encodes a molecular sequence that reflects the composition of the external solution. However, the spacing distributions exhibit a pseudo-Gaussian profile (Fig S2), deviating markedly from the exponential statistics expected for a one-dimensional Bernoulli insertion process. This behaviour suggests that entering molecules experience attractive or repulsive interactions at the BNNT gate, perturbing a purely Poissonian loading mechanism. In addition, the finite BNNT length (1–3 µm) limits the observation of large spacing events, while the optical resolution (~250 nm) prevents resolving shorter distances.

Because achieving sub-wavelength separations of emitters while minimizing their aggregation is essential for enhancing coherent inter-emitter coupling, we next evaluate the gating efficiency of the BNNT pore at higher 6T content, focusing on $R_{1:5}$ and $R_{1:1}$. To directly probe the chemical composition of individual EQC@BNNTs bellow the diffraction limit, we performed aberration-corrected scanning transmission electron microscopy combined with electron energy-loss spectroscopy (ac-STEM-EELS) at 80 kV on

EQC@BNNTs samples spanning compositions from pure 6T ($R_{1:0}$) to pure anthracene ($R_{0:1}$), with intermediate ratios $R_{1:1}$, $R_{1:5}$. The EQC@BNNT were suspended in vaccum on TEM grid by drop-casting and drying. The spatial resolution of our STEM-EELS apparatus is 1 nm. (Fig. 1e ; see Supplementary Information for details). Element-resolved EELS maps acquired at the boron (192 eV), carbon (285 eV) and sulfur (165 eV) edges reveal a clear evolution of the molecular arrangement as a function of $R_{6T:Anth}$. Increasing the 6T content progressively reduces the spacing between sulfur-rich regions, corresponding to 6T emitters, from quasi-continuous chains to discrete emitter sequences separated at the nanometer scale.

Absorption spectroscopy of post-synthesized and filtered EQC@BNNT dispersions provides an independent probe of the overall 6T loading as a function of $R_{6T:Anth}$ (Fig. 1g). EQC@BNNTs filled exclusively with anthracene ($R_{0:1}$) exhibit a single absorption band centred at 350 nm. (Fig. 1g) Progressive incorporation of 6T in the precursor solution results in the emergence and growth of a distinct absorption band at 530 nm, characteristic of encapsulated 6T molecules in BNNT[20,22]. The integrated area of this band scales linearly with the 6T fraction over more than three orders of magnitude (Fig 1h), demonstrating quantitative and robust control of emitter loading through the synthesis ratio.

Upon normalization to the 6T vibronic transition at 485 nm, the absorption spectra reveal a relative enhancement of the 0-0 and 0-1 vibronic features at the highest 6T ratios (Fig 1i). This spectral evolution is a hallmark of molecular interaction associated with head-to-tail J-type aggregation of the optically bright excitonic state.[24,25] The effect is most pronounced for EQC@BNNTs exhibiting the shortest intermolecular spacings, consistent with enhanced excitonic coupling at high packing densities. Together, the structural and spectroscopic analyses presented in Fig. 1 establish dual encapsulation of emitters and spacers as a robust strategy to statistically encode a tailored one-dimensional dielectric function, with controllable spacing distances spanning from 1 nm to micrometres. This level of spatial control defines a scalable platform for engineering and tuning interactions between quantum emitters along a single chain. These results motivated a detailed investigation of the fluorescence properties of 6T emitters versus $R_{6T:Anth}$.

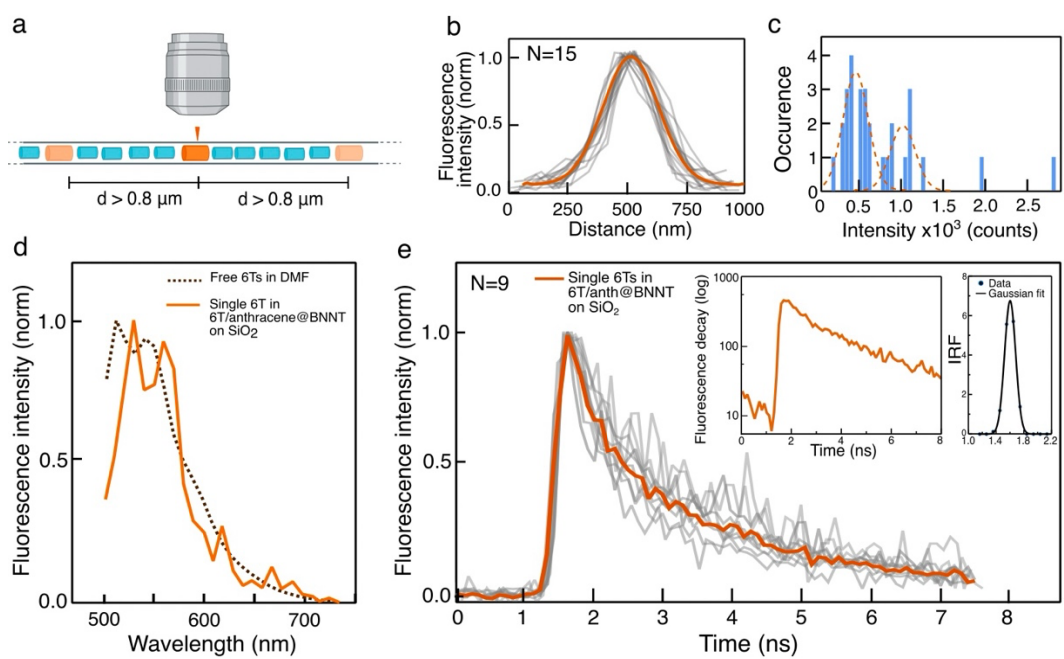

**Figure 2 - Optical signatures of isolated emitters in Encoded Quantum Chains. a** Selection scheme of fluorescence intensity spots separated by more than 800 nm used for optical measurements, from $R_{1:500}$ $R_{1:250}$. **b** Superposition of fluorescence intensity profiles from individual isolated spots (grey) and their averaged profile (orange). **c** Distribution of integrated fluorescence intensities from an ensemble (N=15) of spatially isolated 6T emitters encapsulated in BNNTs. Sub-distributions are highlighted by an orange dashed line. **d** Fluorescence emission spectra of free 6T molecules in DMF and of an individual 6T emitter encapsulated in a BNNT, recorded under 488 nm excitation. **e** Time-resolved fluorescence decays from an ensemble (N=9) of isolated 6T emitters (grey) and their averaged decay (orange), measured under 488 nm excitation. Inset: semi-logarithmic representation of the decay from a single 6T emitter in a BNNT together with the instrument response function (IRF).

From the sample library established beforehand, we first examine EQC@BNNTs displaying widely spaced fluorescence intensity spots, separated by distances exceeding the excitation wavelength and fixed at 800 nm, as illustrated in Fig. 2a. The fluorescence profiles of these spots exhibit similar full widths at half maximum (FWHM) (Fig. 2b), close to the optical resolution limit of the imaging setup. In addition, the integrated fluorescence intensities cluster into two discrete distributions, with a dominant population at $0.4 \times 10^3$, the lower intensity maximum value (Fig. 2b).

These observations point to a homogeneous population of emitters and suggest that the majority of fluorescence spots correspond to single 6T molecules. To test this hypothesis, we compared the fluorescence spectra and excited-state lifetimes, measured upon 488 nm excitation, of isolated 6T molecules in DMF with those recorded from individual fluorescence spots along EQC@BNNTs at room temperature (Fig. 2c,d). The emission spectra are similar, with a slight narrowing of the linewidth observed for 6T molecules encapsulated in BNNTs, likely reflecting dielectric screening and structural stabilization of the elongated molecular backbone along the nanotube wall. The observation of quasi mono-exponential decay dynamics (Fig 2d), together with characteristic lifetimes comparable to those of individual 6T molecules in Poly(methyl methacrylate) (PMMA) matrix environments (1-2 ns)[26], indicates that the fluorescence spots predominantly arise from single molecule emitters.

Having established a well-defined isolated emitter regime at large intermolecular separations, we now investigate the regime of reduced spacing, where interactions between emitters are enhanced. The key parameters governing coupling between identical molecular emitters are their relative dipolar alignment and reciprocal separation. To probe interaction signatures, we capitalize on the versatile assembly platform introduced in Fig. 1, in which the BNNT host enforces a common alignment of the 6T transition dipoles, while anthracene molecules act as adjustable spacers along a single chain.

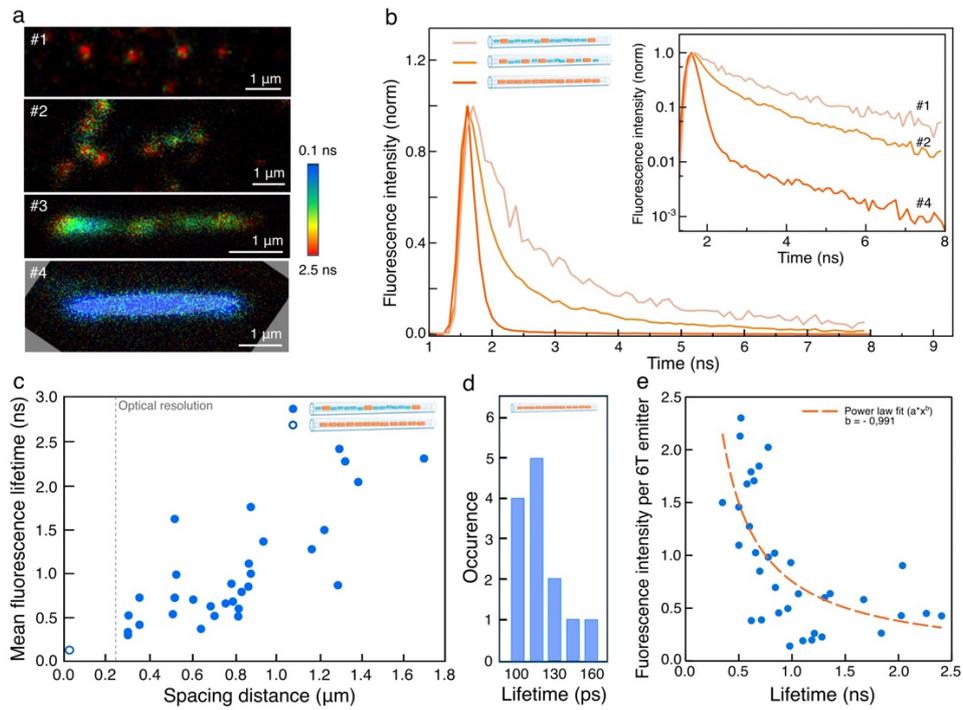

**Figure 3 - Coupling of Molecular Quantum Emitters in BNNTs at room temperature**. **a** Fluorescence Lifetime Imaging (FLIM) at an excitation wavelength of 532 nm of a continuous chain of 6T emitters (#4) and progressively spaced 6T emitters by anthracene molecules (#3-#2-#1). **b** Fluorescence intensity decays recorded from sample #1, #2 and #4 in (a). Inset : log scale representation of the decays. The EQC@BNNT are deposited on Si/SiO$_2$ substrates. **c** Pondered mean effective lifetimes ($\tau_{eff}$) associated with fast and slow decay components ($\tau_1$ and $\tau_2$), attributed to super-radiant and sub-radiant decay channels, respectively, as a function of inter-emitter spacing. **d** Lifetime distribution measured from an statistical ensemble (N=13) of continuous chains of 6T emitters inside BNNTs. **e** Integrated fluorescence intensity of spaced 6T emitters as a function of their lifetime, fitted by a power law.

Figure 3a presents fluorescence lifetime images (FLIM) of individual EQC@BNNTs spanning configurations from micron-scale separated 6T emitters to quasi-continuous chains, labelled #1 to #4. A progressive acceleration of the fluorescence decay exceeding one order of magnitude is observed as the intermolecular spacing decreases, with lifetimes shortening from approximately 2 ns for isolated emitters down to 100 ps for densely packed chains. Intermediate configurations (#2 and #3) exhibit consistently intermediate lifetime values.

This trend is directly reflected in the time-resolved fluorescence decays shown in Fig. 3b for representative configurations (#1, #3 and #4), measured at room temperature. When displayed on a semi-logarithmic scale (Fig. 3b, inset), the decay dynamics deviate markedly from the mono-exponential behavior observed for isolated emitters. As the inter emitter separation is reduced, a fast and intense decay component emerges and progressively dominates the dynamics. The coexistence of a rapid photon burst in the range of 100 ps ($\tau_1$) and a weak long-lived tail with a decay time ($\tau_2$), comparable to that of isolated emitters, is consistent with the simultaneous presence of super-radiant and sub-radiant decay channels in densely packed emitters in EQC@BNNTs. To sum up roughly, superradiance is a quantum, vacuum-seeded cooperative emission process. Vacuum fluctuations at t=0 trigger the first spontaneous dipole response, which selects a collective phase shared by emitters coupled to the same electromagnetic reservoir. Once phase-synchronized within the wavelength-scale coherence window, the ensemble radiates through a bright collective mode whose coupling to the vacuum scales quadratically with the number of coherent emitters, thereby accelerating the emission and shortening the radiative lifetime compared to isolated molecules.

To quantitatively capture these two contributions, we define an effective fluorescence lifetime, $\tau_{eff}$, calculated as the weighted average of $\tau_1$ and $\tau_2$. Figure 3c displays $\tau_{eff}$ extracted from all measured decays as a function of the inter-emitter spacing. A pronounced shortening of $\tau_{eff}$, with a maximum factor of 15, is observed for separations below approximately 0.8 μm, corresponding to distances comparable to the optical wavelength.

In the ideal Dicke limit, where emitters are confined within a sub-wavelength volume, the collective radiative decay rate is expected to vary inversely with the number of coherently coupled emitters. More generally, in three-dimensional geometries, collective emission is governed by the optical coherence volume, leading to a cubic dependence on the inter emitter spacing. In contrast, the present EQC@BNNT system exhibits an approximately linear increase of the effective fluorescence lifetime with the increase of emitter separation. This behavior arises from the one-dimensional confinement imposed by the finite-length nanotube architecture. In this geometry, only emitters located within a segment of the chain of length comparable to the emission wavelength can radiate coherently. As a consequence, the effective number of mutually coherent emitters decreases linearly with the intermolecular spacing along the chain. This geometric argument accounts for the experimentally observed scaling of $\tau_{eff}$. A minimal analytical model capturing this behavior is provided in the Supporting Information.

In the case of continuous emitter chains (#4), the slow decay component $\tau_2$ carries a negligible weight, such that the fluorescence dynamics are dominated by the fast component $\tau_1$. Under these conditions, the effective emission dynamics are well captured by a characteristic timescale of about 100 ps (Fig. 3d). This nearly order-of-magnitude reduction of the fluorescence lifetime compared to the nanosecond-scale decay of isolated emitters indicates the formation of a delocalized radiative channel involving an effective number of on the order of ten coupled emitters at room temperature (see Supporting Information for details).

In Figure 3e, the relation between the effective lifetime $\tau_{\text{eff}}$ and the emission intensity per single 6T emitter is shown. Only spatially resolved emitters are considered. Strikingly, the local emission intensity follows an inverse proportionality to $\tau_{\text{eff}}$, with a power-law exponent $b \approx -1$. This scaling reflects a regime in which the fluorescence brightness is directly proportional to the radiative decay rate, indicating that the accelerated emission arises from enhanced collective radiative coupling rather than from changes in excitation efficiency or non-radiative losses. Consistently, this behavior can be interpreted as a redistribution of oscillator strength among collective radiative channels, where fast, superradiant decay components dominate the detected fluorescence, while long-lived subradiant components contribute only weakly.

The scaling observed in Fig. 3 relies on the one-dimensional nature of isolated EQCs. We therefore examine how the decay evolves when EQCs are brought into close proximity, thereby modifying the spatial arrangement of emitters beyond a single chain. (Fig4a)

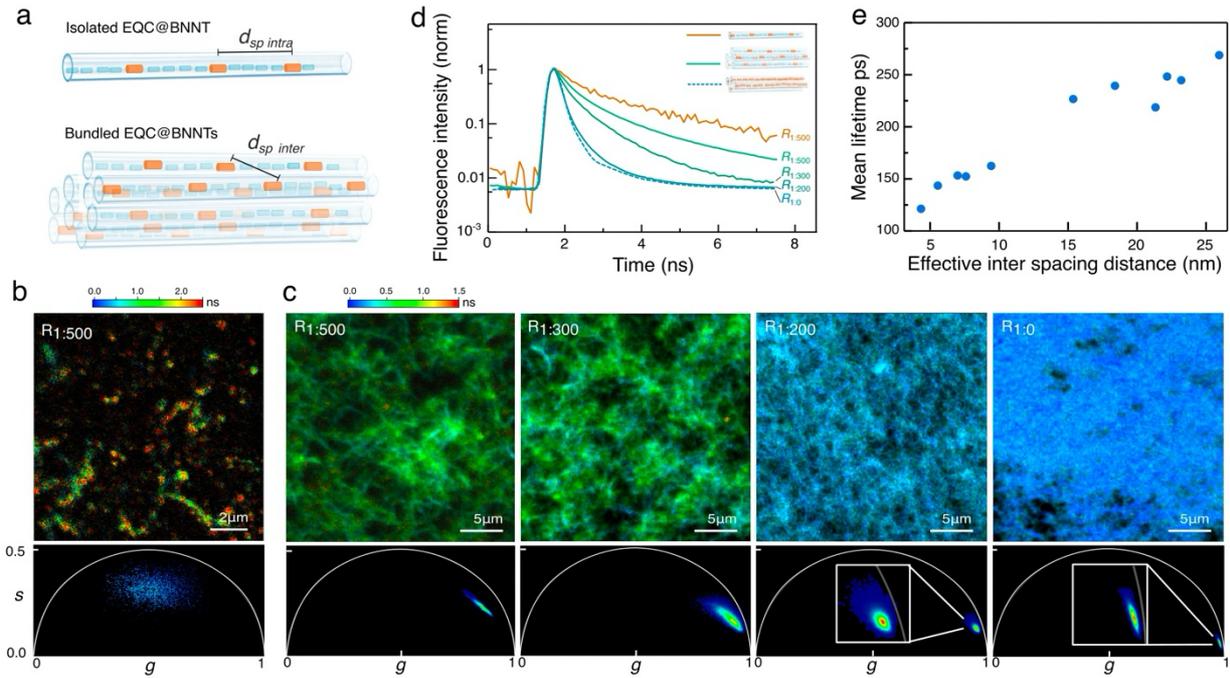

**Figure 4. Dimensional crossover from one-dimensional to higher-dimensional collective emission. a** Schematic illustration of EQC@BNNT bundling at a given $R_{6T:Anth}$ ratio, promoting inter-tube coupling between emitters. **b-c** Fluorescence lifetime imaging microscopy (FLIM) with corresponding PHASOR representation of a sub-monolayer of micro bundles EQC@BNNT (b) and of densely bundled EQC@BNNT mats, deposited on a Si/SiO$_2$ substrate, for the BNNT filling conditions $R_{1:500}$, $R_{1:300}$, $R_{1:200}$ and $R_{1:0}$, under 514 nm excitation. Lifetime maps are obtained from the intensity-weighted mean photon arrival time without fitting. **d** Semi-logarithmic representation of the fluorescence decays recorded from isolated and bundled EQC@BNNTs synthesized at $R_{1:500}$, $R_{1:300}$, $R_{1:200}$ and $R_{1:0}$, extracted from the regions imaged in (b-c). **e,** Evolution of the mean fluorescence lifetime extracted from the fluorescence decay fitting of bundled EQC@BNNT synthesized at different $R_{6T:Anth}$ ratios, plotted as a function of the calculated effective inter emitter spacing within the bundle.

To probe cooperative emission beyond a single one-dimensional chain, EQC@BNNTs were drop-cast onto a Si/SiO$_2$ substrate and slowly dried in an oven at 50 °C. This procedure promotes van der Waals–driven adsorption and spontaneous bundling of BNNTs into assemblies with typical lateral dimensions of a few hundred nanometres, which further aggregate into extended mats on the substrate.

Figures 4b–c present fluorescence lifetime imaging microscopy (FLIM) maps together with their corresponding phasor representations[27,28] of EQC@BNNT assemblies prepared at increasing BNNT filling conditions ($R_{1:500}$, $R_{1:300}$, $R_{1:200}$ and $R_{1:0}$).

At sub-monolayer coverage (Fig. 4b), where EQC@BNNTs are partially isolated from one another, the FLIM images reveal pronounced spatial variations of the fluorescence lifetime. This heterogeneity is reflected in broad phasor distributions, indicating a wide range of local BNNT densities and radiative decay channels, despite identical synthesis conditions. In contrast, EQC@BNNTs synthesized at the same $R_{1:500}$ ratio but deposited as densely bundled mats exhibit reduced and spatially homogeneous fluorescence lifetimes across the $15 \times 15$ μm$^2$ imaged areas.

This constitutes a direct signature of inter-chain coupling: while isolated EQC@BNNTs retain lifetimes comparable to those measured in individual nanotubes, bundling alone induces a pronounced shortening of the lifetime, despite identical emitter spacing within the BNNTs. This observation provides direct evidence that collective radiative interactions extend beyond a single one-dimensional chain and arise from coupling between emitters hosted in neighbouring BNNTs within the bundle.

Mats of bundled EQC@BNNTs synthesized at various $R_{6T:Anth}$ ratios were further investigated by time-resolved imaging (Fig. 4c), revealing a homogeneous lifetime shortening as the emitter density is increased. The corresponding phasor distributions progressively shift toward shorter lifetimes and become increasingly narrow with increasing 6T loading, indicating a homogenization of the radiative dynamics within the bundles. The mean lifetime extracted from these densely packed assemblies decreases from approximately 0.7 ns at $R_{1:500}$ to about 0.1 ns at $R_{1:0}$.

Consistently, the fluorescence decays extracted from the regions highlighted in Fig. 4b–c exhibit a progressive acceleration with increasing emitter density (Fig. 4d). In the dense limit ($R_{1:0}$), corresponding to quasi-continuous emitter chains assembled into tightly packed bundles, the decay becomes strongly multi-exponential, reflecting the coexistence of multiple collective radiative channels associated with heterogeneous local coupling configurations.

Building on the evidence of inter-EQC coupling in bundled configurations, we estimated the effective inter-emitter spacing within BNNT bundles by combining the experimentally determined linear emitter densities with geometrical packing considerations (see Supplementary Information). This enables a direct correlation between the mean fluorescence lifetime of bundled EQC@BNNTs and the effective three-dimensional emitter separation for each encapsulation ratio (Fig. 4e). A key observation is that collective emission in bundled architectures evolves over a markedly reduced length scale compared to the strictly one-dimensional case. In bundles, a substantial reduction of the mean lifetime—from approximately 300 ps to 100 ps—occurs for a change in effective inter-emitter spacing of only ~25 nm. By contrast, in isolated one-dimensional EQCs, a comparable lifetime variation requires an increase in spacing of at least ~250 nm (Fig. 3c).

This pronounced difference highlights a dimensional crossover in the collective radiative dynamics: while emission in isolated EQCs is governed by intra-chain coupling along a single one-dimensional axis, BNNT bundling activates additional radiative pathways mediated by inter-chain interactions. As a result, collective emission becomes significantly more sensitive to spatial organization, reflecting the emergence of higher-dimensional coupling channels enabled by the bundled geometry.

**Discussion**

This work introduces a general strategy to program collective and cooperative emission through the spatial encoding of molecular emitters in confined geometries. Rather than relying on deterministic positioning, interactions are controlled statistically via the co-encapsulation of emitters and spacer molecules within a one-dimensional dielectric host. This approach defines Encoded Quantum Chains (EQCs) as a chemical route to engineer radiative coupling by design.

While demonstrated here with organic emitters confined in boron nitride nanotubes, the EQC concept is not restricted to a specific molecular system. In principle, any combination of optically active nano-objects and inert spacer molecules compatible with one-dimensional confinement can be incorporated, enabling the modular assembly of tailored emitter sequences. In this sense, Encoded Quantum Chains provide a molecular building-block platform in which interaction strength, emitter density, and effective dimensionality are encoded at the synthesis stage rather than imposed post-fabrication.

The continuous evolution observed from isolated single emitters to strongly interacting ensembles, and further to inter-chain coupling through nanotube bundling, reveals how collective radiative behaviour progressively emerges as a function of spatial organization. Importantly, this evolution is accessed within a single material platform and under ambient conditions, highlighting both the robustness of the approach and its relevance for scalable photonic architectures based on molecular emitters.

Beyond the specific case of one-dimensional chains, EQCs offer a versatile experimental framework to revisit the photophysics of zero-dimensional emitters, such as molecular quantum dots or nanocrystals, when they are brought into controlled interaction with one another or with a structured electromagnetic environment. By bridging the gap between isolated emitters and collective ensembles, this platform enables systematic exploration of how radiative rates, coherence, and emission statistics evolve from the single-particle limit to interacting many-body regimes.

An intriguing aspect of our observations is that signatures of coupling persist at intermolecular separations approaching several hundred nanometres, well beyond the molecular scale, and in a condensed dielectric environment. While such distances remain compatible with one-dimensional radiative coupling within an optical wavelength, the robustness of the effect suggests that the underlying physical picture may not reduce to a simple ensemble of dipoles coupled through free-space modes. In particular, the boron nitride nanotube host itself may contribute to stabilizing or mediating delocalized electromagnetic modes along the chain, despite its wide-bandgap dielectric nature[19]. Contributions from defect-related states, modified local photonic density of states, phononic interactions[29], or weakly guided optical modes supported by the nanotube geometry cannot be excluded at this stage. Similar environmental effects are known to strongly influence radiative dynamics and coherence in other low-dimensional systems, such as two-dimensional semiconductors encapsulated in hexagonal boron nitride[30,31]. Elucidating the interplay between molecular excitons, the nanotube host, and their shared electromagnetic environment therefore represents an important direction for future work.

More broadly, Encoded Quantum Chains establish a robust and manipulable platform to explore collective light–matter interactions in low-dimensional molecular systems, spanning regimes from isolated single emitters to superradiant ensembles and higher-dimensional coupled networks. Because EQCs rely on a solid-state nano-template that can be processed in solution, deposited on surfaces, bundled, or embedded within polymer matrices, they are naturally compatible with integration into optoelectronic and photonic devices. By enabling programmable control over emitter spacing, interaction strength, and dimensionality at the molecular scale, this approach opens new avenues for engineering delocalized radiative states, distributed quantum emitters, and emergent many-body optical phenomena in scalable solid-state architectures.

**Acknowledgments**

The authors warmly acknowledge C. Kingston and his team at the National Research Council Canada for donation of BNNTs materials. E.G. acknowledges funding from CNRS starting package and CNRS Tremplin program, the GDRi Howdi. E.G. is supported by the NAQUIDIS Quantum Center and Region Nouvelle Aquitaine.  E. G. and G. R. are supported by the CNRS MITI 'Défi Auto-Organisation' grant. G. R. acknowledges the GdR ImaBio for support, and the ANR for funding (ANR-21-CE45-0028). Confocal fluorescence microscopy and lifetime imaging were performed on the Bordeaux Imaging Center a service unit of the CNRS-INSERM and Bordeaux University, member of the national infrastructure France BioImaging supported by the French National Research Agency (ANR-10-INBS-04). This work has benefited from the Electronic Microscopy MOSTRA platform of the Laboratoire d'Etude des Microstructures (LEM UMR 104 CNRS-ONERA), and the financial support of Région Île-de-France within the project SESAME 2020. The authors acknowledge C. Salomon and A. Browaeys for discussion on superradiance.

**Authors Contributions**

J.-B.M., and E.G. designed the experiments, J.-B. M. and J. L.B. prepared the samples, J.-BM, J. L.B., C.P., G.R., E.G. performed the optical experiments and analysed the results. J.L.B, F.F. and A.L. performed and discussed the STEM-EELS experiments. All authors contributed to the scientific discussions, manuscript preparation and final version.

**Competing interests**

The authors declare no competing interests

Supporting information file for

# Networking Quantum Emitters on a Single Chain
From Single to Cooperative Emitters

Jean-Baptiste Marceau[1#], J. Le Balle[1,2#], Christel Poujol[3], Frederic Fossard[2], Annick Loiseau[2], Gaëlle Recher[1] and Etienne Gaufrès[1*]

[1] Laboratoire Photonique Numérique et Nanosciences, Institut d'Optique, CNRS UMR5298, Université de Bordeaux, F-33400 Talence, France

[2] Laboratoire d'Étude des Microstructures, ONERA-CNRS, UMR104, Université Paris-Saclay, BP 72, 92322 Châtillon Cedex, France

[3] Bordeaux Imaging Center, UAR 3420 CNRS – Université Bordeaux – US4 INSERM, Centre Broca Nouvelle Aquitaine, 33076 Bordeaux, France

\# the authors contributed equally

*Correspondence: etienne.gaufres@cnrs.fr

# 1. EQC@BNNT synthesis

In order to achieve the different encapsulation ratios, we start by preparing the stock solutions of α-sexithiophene (6T) and anthracene (Anth). The 6T stock solution is obtained by dissolving 1.98 mg of α-sexithiophene powder from Sigma-Aldrich in a volume V=500 mL of N,N-dimethylformamide (DMF), yielding a final concentration of $C = 0.8 \times 10^{-5}$ mol/L. To simplify the ratio calculations, we prepare the anthracene stock solution at the same concentration as 6T. Therefore, we dissolve 1.14 mg of anthracene powder from Sigma-Aldrich in a volume V=800 mL of toluene.

In order to prepare the 6T-Anth ratios, we decided to produce a final volume V=100mL. The table below shows the aliquot volumes taken from the stock solutions of 6T and Anth in order to obtain the different ratio solutions.

| Ratio (6T:Anth) | 6T stock solution pipetted | Anth stock solution pipetted |
|---|---|---|
| 1:0 | 100 mL | 0 mL |
| 1:1 | 50 mL | 50 mL |
| 1:5 | 20 mL | 80 mL |
| 1:50 | 1,961 mL | 98,04 mL |
| 1:100 | 990,1 µL | 99,0099 mL |
| 1:250 | 400 µL | 99,60 mL |
| 1:500 | 199,6 µL | 99,80004 mL |
| 0:1 | 0 mL | 100 mL |

The encapsulation of the molecules inside the BNNTs is carried out at reflux at 80°C for 24 hours. In order to ensure the repeatability of the experiments we applied the following protocol:

To ensure their homogeneity, the stock solution and the BNNT solution (pre-treated and opened BNNTs) undergo a 5 minutes sonication. 20 mL of the stock ratio solution and 6 mL of the BNNT solution are pipetted and placed in a round-bottom flask with a magnetic stirrer. The round-bottom flask is then connected to the Allihn condenser which is itself connected to the water system. The setup – a crystallization dish with an oil bath placed on a hot plate with magnetic stirring – is raised so that the round-bottom flask is immersed to 50%.

At the end of the 24-hour reaction, the oil bath is lowered in order to retrieve the flask. The solution is then filtered through a PTFE membrane using a Büchner Funnel. The PTFE membrane presents 0.2 µm pores which were previously opened using DMF. The membrane is then transferred in a flask containing DMF and placed in a sonication bath to detach the mol@BNNT from its surface. Finally, the membrane is retrieved from the flask, resulting in our final mol@BNNT solution.

# 2. Bundling EQC@BNNTs on substrate

Si/SiO$_2$ marked wafers are cleaved to obtain approximately 7x7 mm substrates. They are then placed in an acetone bath under sonication for 10 minutes followed by an isopropan-2-ol bath under sonication for 10 minutes, before being dried with nitrogen gas. The substrate is placed on a piece of optical paper positioned on a glass slide. The solution of the desired mol@BNNT ratio is deposited onto the substrates' surface by drop-cast deposition. In order for the solvent to evaporate, the sample is placed in an oven at 70°C for 2 hours.

## 3. Structural and chemical analysis of the EQC@BNNT using STEM EELS

The studied Mol@BNNT solutions are deposited onto Lacey carbon 400-mesh copper grids and analysed with Scanning Transmission Electron Microscopy – Electron Energy Loss Spectroscopy (STEM-EELS) on S/TEM MOSTRA Microscope (JEOL NeoARM) at 80kV. As we are analysing a complex sample, it is important to choose an energy range that allows us to record the sulphur, boron and carbon signatures, respectively 165 eV, 192 eV and 285 eV, therefore we centre at 200eV. Once the measurement is achieved, the data is analysed with the Panta Rhei software. This software generates an interactive elemental EELS image by selecting the elemental core value of the element we wish to study. By creating different maps for the different elements, it is possible to create an overlay of the different contributions.

## 4. Confocal Fluorescence Lifetime measurements

<u>Setup apparatus</u> The samples prepared in (2) are used for Fluorescence Lifetime measurements. The ladder were performed on a confocal microscope model Leica microscope SP8 WLL2 on an inverted stand DMI6000 (Leica Microsystems, Mannheim, Germany) using a HCX Plan Apo CS2 63X oil NA 1.40 objective. Since the microscope is inverted and our samples are not transparent we must prepare them in a way such as to physically invert them. To achieve this, the sample is placed on a glass slide in a way that the side to be studied is facing upwards. Drops of superglue are placed on either side of the sample. A drop of glycerol is placed on the surface of the sample before covering it with a glass coverslip. The coverslip will hold in place thanks to the superglue. This setup allows the sample to be inverted while having the necessary dimensions for the microscope. The fluorescence lifetime and confocal imaging measurements are carried out with an excitation wavelength of 514 nm. The emission is collected with a hybrid detector ranging from 500 nm to 700 nm. The data is then analysed through FLIMX software, enabling us to obtain the different lifetime values, fit and phasor diagram.

<u>Measurement of inter emitter spacing along a EQC@BNNT</u>. The spacing distance was defined as the distance between the centroids of fluorescence dots as highlighted on Figure S1. Measurement of the distance was done directly using the FLIMX software.

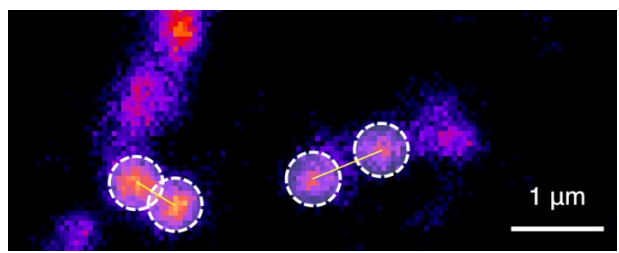

**Figure S1** - Determination of inter emitter spacing based on dots centroids

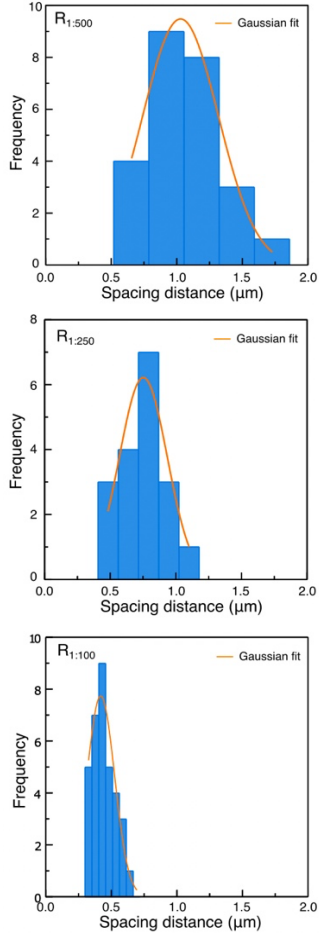

**Figure S2** Spacing distance distribution and Gaussian fit as a function of $R_{6T:Anth}$ ($R_{1:500}$, $R_{1:250}$, $R_{1:100}$) measured from confocal images recorded at an excitation wavelength of 514 nm and a collection range of 550-700 nm.

## 5. Analytical model for 1D collective decay in a finite nanotube

### 5.1. Periodic chain with constant spacing

**Geometry and assumptions.**

We consider a finite, quasi-one-dimensional chain of $N$ identical two-level emitters confined inside a BNNT. Emitters are equally spaced by $a$ at axial positions $z_i = i\,a$ ($i = 1 \dots N$), with single-emitter radiative decay rate $\Gamma_0$ and transition wavelength $\lambda$ ($k = 2\pi/\lambda$). The BNNT enforces a predominant alignment of transition dipoles with the tube axis, so the axial projection of the electromagnetic Green tensor governs the dissipative (radiative) coupling.

**Coupling kernel and collective rate.**
The radiative coupling between emitters $i$ and $j$ reads

$$\Gamma_{ij} = \begin{cases} \Gamma_0, & i = j \\ \Gamma_0 \dfrac{\sin(k\, r_{ij})}{k\, r_{ij}}, & i \neq j \end{cases}$$

with $r_{ij} = |i - j|\, a$.

For a collective eigenmode $\mathbf{c} = (c_1, \ldots, c_N)$, the total radiative decay rate is

$$\Gamma[\mathbf{c}] = \sum_{i,j=1}^{N} \Gamma_{ij}\, c_i c_j^*.$$

Because the phase factor $\sin(kr)/(kr)$ oscillates beyond $r \sim \lambda$, only emitters within an axial window of size $\sim \lambda$ add constructively. The effective coherent size of the bright (superradiant-like) mode is therefore

$$N_{\text{eff}}^{(0)} \sim \frac{\lambda}{a},$$

leading to the collective decay rate

$$\Gamma_{\text{coll}}^{(0)} \approx \Gamma_0\, N_{\text{eff}}^{(0)} \sim \Gamma_0\, \frac{\lambda}{a},$$

and the linear 1D scaling for the bright-mode lifetime

$$\tau_1 \sim \frac{1}{\Gamma_{\text{coll}}^{(0)}} \propto a.$$

### 5.2. Effect of pseudo-Gaussian spacing disorder

In the experiments, the inter-emitter spacing for a fixed loading ratio is not perfectly periodic but follows a pseudo-Gaussian distribution (Fig 1 and Fig S2) around a mean $\bar{a}$. To account for this, we model the axial positions as

$$z_i = i\, \bar{a} + \epsilon_i,$$

Where $\bar{a}$ is the mean nearest-neighbor spacing and $\epsilon_i$ is the zero-mean positional deviation of site $i$ relative to the periodic lattice $i\, \bar{a}$ (we set $\epsilon_1 = 0$ without loss of generality). The local spacing fluctuation (nearest-neighbor) is

$$\delta_i \equiv (z_{i+1} - z_i) - \bar{a},$$

so that

$$\delta_i = \epsilon_{i+1} - \epsilon_i \Leftrightarrow \epsilon_i = \sum_{n=1}^{i-1} \delta_n \quad \text{with } (\epsilon_1 = 0).$$

We assume

$$\langle \delta_i \rangle = 0 \quad \text{and} \quad \text{rms}(\delta_i) = \sigma_a,$$

where $\langle \cdot \rangle$ denotes an ensemble average within the dataset, and $\text{rms}(\cdot)$ is the root-mean-square (standard deviation for zero-mean variables). Thus, $\sigma_a$ is the rms nearest-neighbor spacing fluctuation.

For a pair $(i, j)$, the axial separation entering the coupling kernel is

$$r_{ij} = |z_j - z_i| = |i - j|\bar{a} + \Delta_{ij},$$
$$\text{with}$$
$$\Delta_{ij} \equiv \epsilon_j - \epsilon_i = \sum_{n=i}^{j-1} \delta_n,$$

Under short-range, approximately Gaussian spacing noise (consistent with the pseudo-Gaussian histograms), $\Delta_{ij}$ is also approximately Gaussian with zero mean. Over the wavelength-limited coherent window $|i - j| \lesssim \lambda/\bar{a}$, the rms pairwise jitter is

$$\sigma_\Delta^2 \equiv \langle \Delta_{ij}^2 \rangle \approx \left(\frac{\lambda}{\bar{a}}\right) \sigma_a^2,$$

which controls the statistical phase jitter $k\,\Delta_{ij}$ in the kernel. This yields the Gaussian coherence dephasing factor

$$\mathcal{C} \equiv \langle e^{ik\Delta_{ij}} \rangle \simeq \exp\left[-\tfrac{1}{2} k^2 \sigma_\Delta^2\right] = \exp\left[-\tfrac{1}{2}(2\pi\,\sigma_\Delta/\lambda)^2\right].$$

**Collective rate with spacing disorder**

The axial coherence window $\sim \lambda$ is unchanged; however, the bright-mode amplitude is attenuated by $\mathcal{C} \in (0,1]$. Consequently, the effective coherent size and collective rate become

$$N_{\text{eff}} \approx \left(\frac{\lambda}{\bar{a}}\right)\mathcal{C},$$
$$\Gamma_{\text{coll}} \approx \Gamma_0\,N_{\text{eff}} \approx \Gamma_0\left(\frac{\lambda}{\bar{a}}\right)\mathcal{C},$$

and the fast lifetime scales as

$$\tau_1 \approx \frac{1}{\Gamma_{\text{coll}}} \propto \frac{\bar{a}}{\mathcal{C}}.$$

The functional form of the 1D trend is unchanged, $\tau_1$ remains linear in the mean spacing $\bar{a}$, but the slope is renormalized by $\mathcal{C} < 1$, which depends on the rms spacing fluctuations through $\sigma_\Delta$.

Practically, larger spacing dispersion (larger $\sigma_a$, hence larger $\sigma_\Delta$) reduces the collective enhancement (smaller $N_{\text{eff}}$), yields longer $\tau_1$ at fixed $\bar{a}$, and broadens the distribution of fitted lifetimes across nominally similar samples. These statements are consistent with the pseudo-Gaussian spacing statistics observed at fixed ratios.

**Contrast with 3D ensembles.**

The disorder correction described above acts primarily on the prefactor of the 1D collective rate. The scaling contrast with 3D ensembles remains: in 3D, $N_{\text{eff}}^{(3D)} \sim (\lambda/\bar{a})^3$ and $\Gamma_{\text{coll}}^{(3D)} \sim \Gamma_0 (\lambda/\bar{a})^3$, i.e. a cubic dependence of lifetime on $\bar{a}$, distinct from the linear 1D behaviour observed here.

## 5. Estimation of the coupled emitter effective number

**Effective number of coherently coupled emitters**

The fluorescence decays of EQC@BNNTs are well described by a biexponential function,

$$I(t) = A_1 e^{-t/\tau_1} + A_2 e^{-t/\tau_2},$$

where $\tau_1$ corresponds to a fast radiative channel and $\tau_2$ to a slower one. In the context of collective emission, the fast component $\tau_1$ is attributed to the bright collective mode formed by emitters that remain phase-synchronized within the coherence window ($\sim \lambda$). The slow component $\tau_2$ arises from weakly radiative configurations (partially dephased or independant dipoles) and therefore provides a practical baseline for the single-emitter radiative rate.

Experimentally, $\tau_2$ remains close to the lifetime of isolated emitters, $\tau_0$, indicating that strongly subradiant modes—while present—carry negligible weight in the detected signal. This behaviour is typical in solid-state molecular systems, where deeply subradiant eigenstates exhibit extremely weak radiative amplitudes and are therefore essentially invisible in time-resolved fluorescence.

To quantify the enhancement of the radiative rate associated with the bright collective mode, we define an effective collective decay rate.

$$\Gamma_{\text{eff}} = \frac{A_1 \Gamma_1 + A_2 \Gamma_2}{A_1 + A_2}, \quad \Gamma_i = 1/\tau_i$$

In the regime relevant to EQCs, the fast component dominates ($A_2 \ll A_1$), so that

$$\Gamma_{\text{eff}} \simeq \Gamma_1 = 1/\tau_1$$

The effective number of coherently coupled emitters is then defined as

$$N_{\text{eff}} = \frac{\Gamma_{\text{eff}}}{\Gamma_0} \simeq \frac{\tau_0}{\tau_1},$$

where $\Gamma_0 = 1/\tau_0$ is the single-emitter radiative rate inferred from isolated 6T molecules or from the slow component $\tau_2$.

Using typical values obtained in this work ($\tau_1 \approx 100$ ps and $\tau_0 \approx 2.0$ ns), we estimate

$$N_{\text{eff}} \approx 20,$$

$N_{\text{eff}}$ reflects the number of emitters that can radiate constructively within the axial coherence window imposed by the one-dimensional geometry of the BNNT. This value is not the total number of molecules within a chain, but rather the size of the wavelength-limited collective dipole that dominates the radiative dynamics.